\documentclass{mem}
\usepackage{natbib}
\usepackage{txfonts}
\usepackage{balance}
\usepackage{graphicx}
\usepackage{flushend}
\usepackage[breaklinks,pdftex]{hyperref}
\idline{75}{282}

\begin{document}

\title{A search for mid-IR bands of amino acids in the Perseus Molecular Cloud}

\author{
Susana \,Iglesias-Groth\inst{1,2} 
}

\institute{
Instituto de Astrof{\'\i}sica de Canarias, E-38200 La Laguna, Tenerife, Spain
\and 
Universidad de La Laguna, Dept. Astrof{\'\i}sica, E-38206 La Laguna, Tenerife, Spain
\email{sigroth@iac.es}}

\authorrunning{Susana Iglesias-Groth}

\titlerunning{amino acids star-forming regions}

\date{Received: 5 december 2022}

\abstract{Amino acids are building-blocks of proteins, basic constituents of all organisms and essential  to life on Earth. They are present in carbonaceous chondrite meteorites and  comets, but their origin is still unknown. We present Spitzer spectroscopic observations in the star-forming region IC 348 of the Perseus Molecular Cloud showing the possible detections of mid-IR emission lines consistent with the most intense laboratory  bands of the three aromatic amino acids, tyrosine, phenylalanine and tryptophan  and  the aliphatic amino acids isoleucine and glycine.  Based on these tentative identifications, preliminary estimates of  column densities give values 10-100 times higher for isoleucine and glycine than for the aromatic amino acids as in some meteorites.  Potential counterparts of the strongest  laboratory bands of each amino acid are also found in the combined  spectrum of  32 interstellar locations obtained in diverse unrelated star-forming regions.
}

\maketitle{}

\section{Introduction}

Proteinogenic and non-proteinogenic amino acids have long been known to be present in meteorites, especially carbonaceous chondrites and their abundances have been compiled in numerous studies \citep{Pizzarello17}, \citep{Koga17}. The presence of amino acids in meteorites and comets \citep{Elsila09} suggests that the young planet Earth was enriched with these molecules, basic constituents of proteins, essential for the emergence of life and the development of living organisms \citep{Botta02}. 

A few dozen molecules have been detected in the interstellar medium of star-forming regions \citep{Favre18} and in protoplanetary disks. These detections include aldehydes, acids, ketones, and sugars. The simplest organic acid, formic acid (HCOOH), which contains the carboxyl group, one of the main functional groups of amino acids, has  been detected in low-mass star-forming regions \citep{Lefloch17} and in a protoplanetary disk \citep{Favre18}. The reaction of protonated alcohols and HCOOH could lead to the production of glycine (and other amino acids) in the hot nuclei of molecular clouds  and formation of amino acids also seems  possible via specific gas-phase reactions in dark clouds \citep{Ehrenfreund00,Elsila07,Redondo17}.   Very sensitive searches of amino acid bands conducted at millimeter wavelengths have focused on glycine, the most simple amino acid, and obtained upper limits to the column density (e.g. \citealt{Ceccarelli00}) or  tentative assignments of ro-vibrational bands \citep{Kuan03} which have not been fully confirmed \citep{Cunningham07}. The young stellar cluster IC 348 (age $\sim$ 2 Myr), located at the eastern end of the well known Perseus Molecular Cloud complex, at a distance of 321 $\pm$10 pc, is one of the nearest star-forming regions  well suited to explore the presence of  complex carbon-based molecules \citep{Luhman16,I-G19}. Organic molecules appear to be widely distributed in this nearby molecular cloud complex which prompted us to carry out an extensive search for amino acids.

Systematic laboratory work \citep{Matei05,I-C18} has led to the identification of a large number of relatively intense amino acid  bands in the mid-IR. Between 15 and 20.8 $\mu$m  these bands are mostly due to the carboxylate group COO- rocking, bending and wagging modes; in the range  20.8 to 26.3 $\mu$m  are mainly associated to the protonated amino group and,  at longer wavelengths, between 26.3 and 37.0 $\mu$m  the bands  are mainly related  to the C-C $\alpha$-N deformation modes \citep{I-C18}.   The Spitzer  Space Telescope  has obtained a large, very valuable number of moderately high resolution spectra of star-forming regions in the spectral range 10-35 $\mu$m. This spectral range appears particularly interesting for a potential identification of individual amino acid species \citep{I-C18}. Among all the 20 proteinogenic amino acids, 
precise laboratory measurements of wavelengths, molar extinction coefficients and integrated molar absorptivities of mid-IR bands are available for the aromatic amino acids tryptophan, tyrosine and phenylalanine and for the aliphatic isoleucine and glycine \citep{I-C22}. In this work, we present results on the search  in the Perseus Molecular Complex for mid-IR bands  (10-35 $\mu$m range) of these five amino acids. 

\begin{figure*}[t!]
\resizebox{\hsize}{!}{\includegraphics[clip=true]{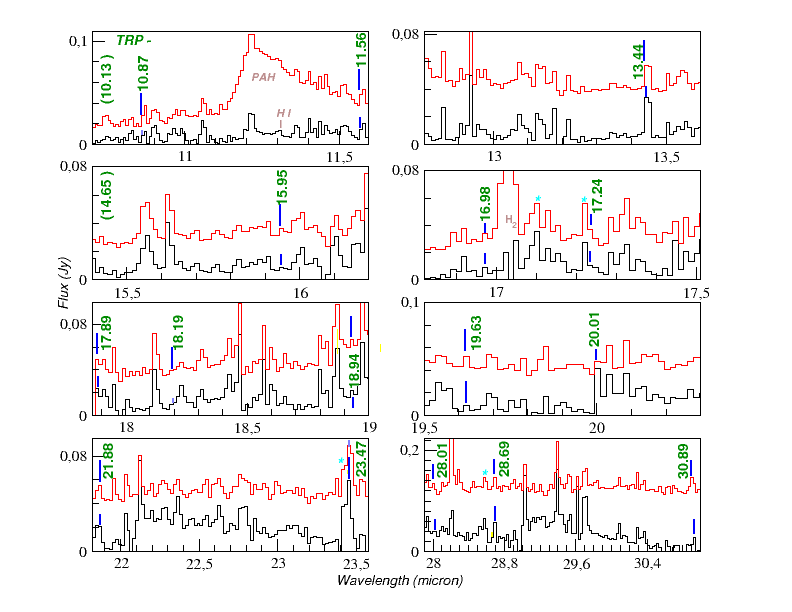}}
\caption{\footnotesize
Tryptophan  (Trp) bands:   laboratory wavelengths  are marked with vertical lines. Emission features assigned  to these bands are indicated  in the combined IC 348 ISM spectrum (red colour) and in the combined star-forming region ISM spectrum (black).  Blue asterisks mark the position of strong water bands. The spectra are shifted on the vertical axis for  convenience of display, only band fluxes relative to the local continuum can be deduced from these plots.
}
\label{fig:Trp}
\end{figure*}

\section{Observations} 
It was used Spitzer Space Telescope moderately high spectral resolution (R$\sim$ 600)  archive spectra  obtained at various interstellar locations  in the central region of  IC 348 in the Perseus Molecular Cloud. All selected slits were  within 10 arcmin of the most luminous star of the  IC 348  cluster:  HD 281159.  Fully reduced   spectra acquired with the Short-High  (S-H, 9.8-19.5 $\mu$m ) and  Long-High (L-H, 19.5-36 $\mu$m)  modules  were taken from the Combined Atlas of Spitzer/IRS Sources, (CASSIS; http:// cassis.sirtf.com, \citep{Lebouteiller15}: provides reduced and flux calibrated data.)
The  individual CASSIS full-aperture extraction  of the  S-H and L-H spectra were subsequently averaged to  produce a high signal to noise spectrum designated hereafter as combined IC 348 ISM.  Similarly, we produced a star-forming region ISM “reference” spectrum by averaging  32 spectra selected from ISM observations of  diverse star-forming regions unrelated to Perseus. Many weak spectral features are found in common which supports they are reliable detections. The minimum flux level  for a 3$\sigma$ detection of  a line in the combined IC 348 ISM  spectrum  is  0.5 x 10$^{18}$ Wm$^{-2}$.  
\begin{figure*}[t!]
\resizebox{12cm}{!}{\includegraphics[clip=true]{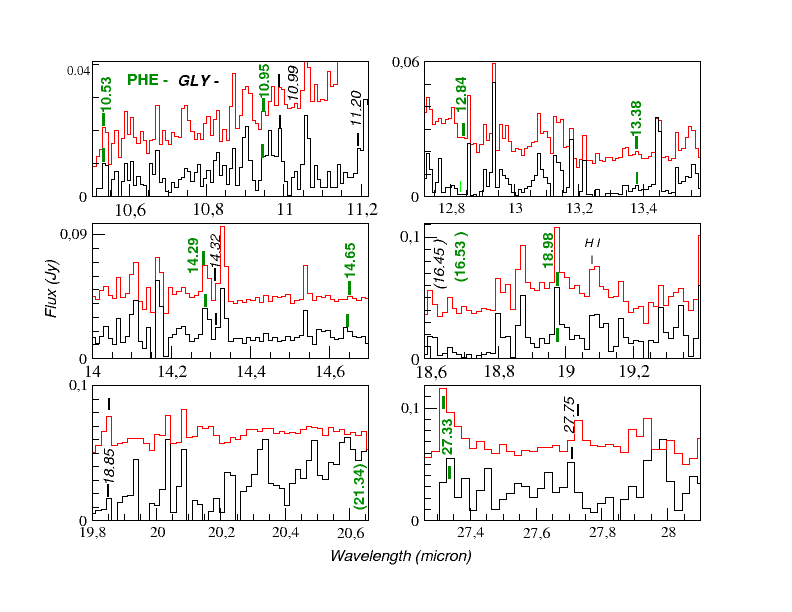}}
\resizebox{12cm}{!}{\includegraphics[clip=true]{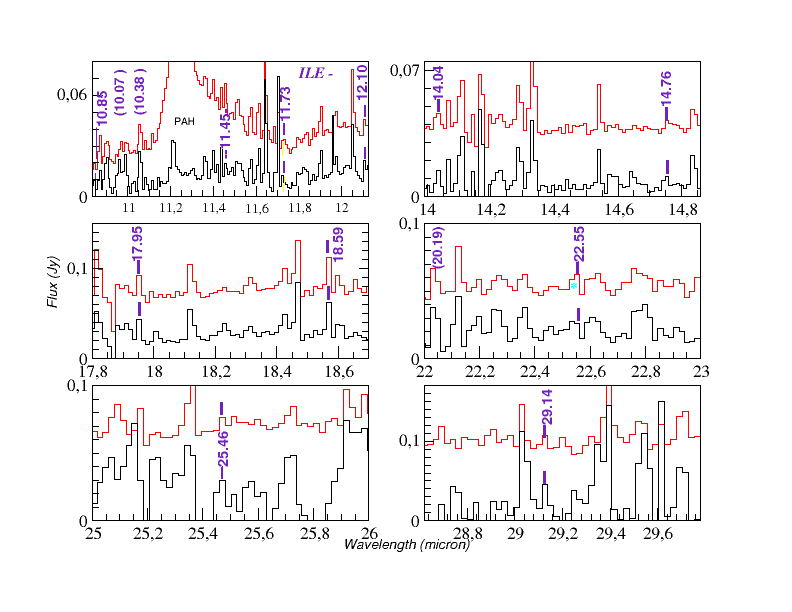}}
\caption{\footnotesize
{\it Top panels}:
Phenylalanine (Phe)  and Glycine (Gly)  bands: caption as in Fig. 1.
{\it Bottom panels}:
Isoleucine (Ile) bands: caption as in Fig. 1.
}
\label{fig:Phe_Ile}
\end{figure*}

\section{Searching for known molecules}

At mid-IR wavelengths there are fundamental vibrational transitions of many relevant molecules which could limit the detectability of amino acid bands.
The HITRAN molecular spectroscopic database was used to explore the presence of these molecular bands in the observed spectrum.  The HITRAN Application Programming Interface (HAPI) was used for downloading and selecting spectroscopic transitions for each molecular specie in the HITRAN online web server and for calculating the corresponding absorption coefficients and absorption spectra in the spectral range of interest (10-34 $\mu$m).  The computed spectra, as well as molecular mid-IR bands reported in the literature served as a reference for their identification in the spectrum of the interstellar gas in  IC 348. In addition, mid-IR transitions of atomic species commonly detected in interstellar environments were considered in this search.
                                  
\section{Results}

Both, the  IC 348  and the reference ISM spectra show  a very large number of emission lines in common. The strongest and broader  feature in the Perseus spectrum is the 11.2 $\mu$m PAH band.   Other strong lines which we can see in the panels of Figures 1-2  are associated to transitions of molecular hydrogen  at 12.28 and 17.04 $\mu$m, and atomic hydrogen at 12.37 and 19.06 $\mu$m. Weaker lines of water, fullerene species and other organic molecules are also  seen in the spectra.  From the excitation diagram of the  rotational transition lines S(0), S(1) and S(2) of molecular hydrogen, we obtain: T$_{ex}$ $\sim$ 270 $\pm$ 10 K and column density N$\sim$ 2.3 x 10$^{21}$ cm $^{-2}$.  It is also found a low abundance of water N $\sim$ 2.7 x 10$^{15}$ cm$^{-2}$, N(H$_2$O/H(H$_{tot}$)$\sim$ 3 x 10$^{-7}$.  
 
Laboratory wavenumbers,  wavelengths (laboratory and observations)  and integrated molar absorptivity ($\Psi$) for the strongest mid-IR transitions of the amino acids under study are listed in  Table 1.  Laboratory wavenumbers  are precise at the level of 1 cm $^{-1}$ and the integrated molar absorptivity adopted from \citet{I-C22}) has uncertainties  of order 20$\%$. In the Figures the positions of the strongest laboratory bands   of each of the five   amino acids  under study  and the  emission features  we tentatively assign to them  are marked on the combined IC 348 ISM  spectrum (red-line). Band assignments are less reliable at  longer wavelengths due to increased  uncertainties in laboratory wavelengths.  For comparison, in each figure it is also plotted the star-forming region  ISM “reference” spectrum (black line). For all  the strongest mid-IR bands of amino acids  we have identified  emission features  with wavelengths consistent within errors with those measured at laboratory. 

Fluxes for the assigned  bands were measured only in the IC 348 ISM spectrum by integrating the continuum subtracted spectrum. In  Table 1 we list, for each of the five  amino acids,  measured fluxes  and wavelengths for the tentative assignments to all the laboratory bands  with  integrated molar absorptivity $\Psi$ $> $ 0.6 km mol $^{-1}$. Errors in the band fluxes were estimated by propagating pixel error through the band integrals and are typically of order 20$\%$. 

TRYPTOPHAN:  In Table 1 there are  listed bands with  reported integrated molar absorptivity $\Psi$ $\geq$ 2 km mol $^{-1}$, we identify in Fig. 1 emission features which  could be associated to each of these laboratory bands and provide fluxes for each of them.   The two  bands with larger integrated molar absorptivity  have  clear counterparts at 13.44 and 19.63 $\mu$m. The next at 11.56 and 23.47 $\mu$m are also clearly detected. For any reported laboratory band with a measured absorptivity we find a counterpart emission feature in the observed spectrum as marked in the figure. However, the weaker bands in the laboratory are not reliably  detected in the observed spectrum. TYROSINE: The strongest absorptivity bands have counterparts at 11.88, 12.59, 15.39, 17.35 and 18.89 $\mu$m but the only non-blended ones are at 11.88 and 15.39 $\mu$m.  This is considered a tentative detection.  PHENYLALANINE: The three stronger laboratory bands have counterparts in the spectrum, other weaker lab. bands have assignments but are less reliable identifications. Overall this may be considered a weak detection.   ISOLEUCINE:  The highest $\Psi$ = 13 km mol $^{-1}$, corresponds to the  18.58 $\mu$m band  which shows the  highest  observed flux in the proposed counterpart emission features. The next bands at 14.05, 22.56, 25.42 and 29.13 $\mu$m  have  also detected counterparts. In some cases, they appear blended, thus it should be considered  a tentative or marginal detection. GLYCINE: There are only six known bands in the spectral range under consideration, out of which five have integrated molar absorptivity reported. The two bands with higher $\Psi$ values at 19.85 and 27.75 $\mu$m and the 14.32 $\mu$m band can be assigned to rather isolated emission lines in the IC 348 spectrum. The band at 10.2 $\mu$m is blended with the strong PAH band and its presence  (and flux) is difficult to ascertain. The remaining band at 16.45 $\mu$m could be associated with an emission line which is however superimposed on  a weak broader spectral feature and the measured  flux  is rather uncertain. The evidence for the presence of glycine  shall be taken with caution.

\begin{table*}
\caption{Laboratory  and observational data: $^m$ \citet{Matei05}, $^r$ \citet{Rosado98};   $^{I-C}$ \citet{I-C21}}
\label{tab:tab1}
\begin{center}
\begin{tabular}{lccccrr}
\hline
\\
Am.Ac & Lab $^{O}$ & Lab$^{I-C}$ & Lab$^{I-C}$ & Obs & $Psi$ & Flux    \\
      & (cm $^{-1}$) &  (cm $^{-1}$) & ($\mu$m) & ($\mu$m) & (km mol $^{-1}$) & (10$^{-18}$ W m$^{-2}$)  \\
\hline
\
\bf{TRP} & - & 987  & 10.13 & 10.14 & 2.3 & 1.4  \\
& -   & 920  & 10.87 & 10.87 & 4.4  & $<$ 2.2  \\
& -   & 865  & 11.56 & 11.57 & 10.8 & 4.5 \\   
& -   & 744  & 13.44 & 14.44 & 30.6 & 6.4 \\
& -   & 683  & 14.64 & 14.66 & 2.3  & 1.4  \\
& 627$^m$ & 627  & 15.94 & 15.95 & 2.7  & 1.4  \\
& 581$^m$ & 588  & 17.01 & 16.98  & -  & $<$ 1.0  \\ 
& 581$^m$ & 580 & 17.24 & 17.24 & 5.6  & 4.5  \\
& 559$^m$ & 559 & 17.89 & 17.89 & 9.8  & 2.8 \\
& 549$^m$ & 550 & 18.18 & 18.19 &  2.8 & 2.6  \\
& 530$^m$ & 528 & 18.94 & 18.93 &  4.0  & 1.5  \\
& 509$^m$ & 509 & 19.64& 19.63 &  31.7 &  3.0 \\
& 499$^m$ & 499 & 20.04 & 20.01 &  4.0  & 1.5 \\
& 456$^m$ & 457 & 21.88 & 21.87 & 1.9  & 2.0  \\  
& 426$^m$  & 426 & 23.47 & 23.47 & 10.0 & 4.8  \\
&  -   & 397 & 25.19 &  25.19 & 0.6 & $<$1.0 \\ 
&  -   & 357 & 28.01 & 28.01 & 1.9 & 2.0  \\
& 347$^m$  & 349 & 28.65 & 28.69 & 9.6  & 3.0  \\
& 325$^m$  & 324 & 30.86 & 30.89 & 3.2 &  3.0  \\
\bf{TYR} & - & 985 & 10.15 & 10.14 & 2.8 & 1.0  \\
& -  & 897 & 11.15 & 11.15 & 2.3 & $<$ 1.2  \\
& -  & 877 & 11.40 & 11.40 & 4.6  & 5.0  \\
& -  & 841 & 11.89 & 11.89 & 14.5 & 3.7 \\
& -  & 794 & 12.59 & 12.59 & 12.7 & 2.2  \\
& -  & 740 & 13.51 & 13.51 & 4.9 & 1.7  \\
& -  & 713 & 14.03 & 14.02 & 1.2 & $<$1.0  \\
& 650$^m$ & 650 & 15.38 & 15.39 & 12.7 & 2.2 \\
& 575$^m$ & 576 & 17.36 & 17.37 & 12.7 & 2.3  \\ 
& 529$^m$ & 530 & 18.87 & 18.87 & 16.0 & 12.0  \\ 
& 493$^m$ & 494 & 20.24 & 20.24 &   2.7 & $<$ 1.5 \\
& 433$^m$ & 434 & 23.04 & 23.06 &  5.5 & 3.0  \\
& 377$^m$ & 380 & 26.32 & 26.33 & 15.6 & 3.4  \\
& 335$^m$ & 335 & 29.85 & 29.72 &  1.3 & 2.2   \\  
\bf{PHE}& - & 950 & 10.53 & 10.53   & 7.2 & 5.7    \\
& -      & 914 & 10.94 & 10.95    & 1.7 & $<$2.0    \\
& -      & 779 & 12.84 & 12.84    & 3.6 & 3.0    \\
& -      & 746 & 13.40 & 13.38    & 9.2 & $<$ 1.8 \\
& -      & 700 & 14.29 & 14.29    & 19.8 & 6.9  \\
& -      & 683 & 14.64 & 14.65    & 2.3  & $<$3.0 \\
& 605$^m$ & 605 & 16.53 & 16.53   &  2.1  & $<2.5$  \\
& 525$^m$ & 526 & 19.01 & 18.98  & 15.0 & 7.2  \\ 
& 469$^m$ & 469 & 21.32 & 21.34   & 5.1 & 3.0 \\
& 365$^m$ & 366 & 27.32 & 27.33  & 54.0 & 10.0  \\ 
\bf{ILE} & -& 993 & 10.07 & 10.07 & 1.3 & $<$ 1.0   \\
& -  & 964 & 10.37 & 10.38 & 1.8 & $<$ 1.3   \\
& -  & 921 & 10.86 & 10.85 & 3.0 & $<$ 2.0   \\
& -  & 873 & 11.45 & 11.45 & 3.2 & 2.5        \\
& -  & 852 & 11.74 & 11.73 & 0.3 & $<$ 1.0 \\ 
& -  & 826 & 12.10 & 12.10 & 0.6 & $<$ 1.0   \\
& -  & 712 & 14.04 & 14.04 & 5.4 & 2.7       \\
& -  & 676 & 14.79 & 14.76 & 3.4 & 2.5       \\ 
& 557$^m$ & 557 & 17.95 &        &   -   &  -  \\ 
& 538$^m$ & 538 & 18.59 & 18.59  & 12.9 & 3.5      \\
& 490$^m$ & 496 & 20.18 & 20.19  & 1.5 & $<$ 1.0  \\ 
& 443$^m$ & 443 & 22.55 & 22.55  & 7.0 & 3.0      \\
& 393$^m$ & 393 & 25.45 & 25.46  & 5.2 & $<$ 1.4  \\
& 343$^m$ & 343 & 29.15 & 29.14  &  4.5 & 3.4       \\
\bf{GLY} & - /911$^r$  & 910  & 10.99 & 10.99 & 10.5 & $<$4.0   \\
& - /893$^r$      & 893  & 11.20 & 11.20 & 10.5 & $<$4.0  \\
& - /698$^r$      & 698  & 14.32 & 14.32 &  6.6 & $<$ 2.5  \\
& 607$^m$/608$^r$  & 608  & 16.45 & 16.45 &  2.1 & $<$1.4  \\
& 504$^m$ / 504$^r$ & 504  & 19.84 & 19.85 & 15.1 & 4.5  \\
 & 356$^m$ / 358$^r$ & 359  & 27.85 & 27.75 & 14.4 & 3.4 \\
\\
\hline
\end{tabular}
\end{center}
\end{table*}

\section{DISCUSSION}

In order to produce vibrational excitation diagrams it is necessary to estimate for each amino acid transition the number of molecules in the upper vibrational state Nu.  This can be obtained from  the  measured  flux of each transition  assuming optically thin emission, since the total power emitted in a band  is  P=N $_u$ A $_{ul}$ h $\nu$$_{ul}$ /(4$\pi$ D $^2$), where, D is the distance to IC 348, $\nu$$_{ul}$ the frequency of the amino acid transition, and h the Planck constant. The Einstein A $_{ul}$ coefficients  can be   obtained from the  laboratory  integrated molar absorptivity.  As molar absorptivity coefficients  are sensitive to temperature and vacuum conditions, we can only expect from the available absorptivity data a very preliminary indication of  the existence of  thermal equilibrium. 

In spite of the limited  laboratory information available, vibration excitation diagrams for each amino acid were obtained  using the strongest transitions for which molar absorptivity were available and  fluxes  measured in the “combined IC 348 ISM” spectrum.  The results show good correlation coefficients in the range r=0.9-0.95 for each individual amino acid, even if for simplicity a vibrational degeneracy g$_u$=1 was adopted for all the energy levels. The inverse of the slopes of the fits  indicates  equilibrium temperatures in the range 270-290 K for each amino acid.   Considering that the  absorbed UV energy is re-emitted mostly via the IR vibrational  bands,  column densities, n(AA), can be derived from the measured band fluxes  in the  IC 348 spectrum.  We will follow procedures similar  to those used to estimate abundances for other molecular species in the ISM  of star-forming regions \citep{Berne17,I-G19}. The total IR intensity (W m$^{-2}$ sr$^{-1}$) emitted by the amino acid AA can be estimated as I $_{tot}$= n(AA) x $\sigma$$_{UV}$ x G $_0$ x 1.2 x 10$^{-7}$ where G $_0$  is the radiation field in the locations of the observations and  $\sigma$$_{UV}$ is the cross section for absorption in the UV of the relevant molecular specie. Large variations exist in the far UV  radiation field  within  the IC 348 star-forming region.  Previous work on fullerene abundances using spectra obtained in very similar locations of the ISM in IC 348, adopted G $_0$ = 45 for the average interstellar radiation field \citep{I-G19}. This value will be adopted here.

The total intensity, I $_{tot}$, can be determined from the  measured total fluxes emitted in the IR. However, we  only measured fluxes for  the subset of stronger lines known in the 10-30 $\mu$ m region. The real total IR intensity will be  higher than the value inferred from our band measurements, and subsequently, the column densities too.  The excitation diagram suggests  that   amino acids  in the ISM of IC 348 are in thermal equilibrium  with excitation temperatures in the range 270-290 K.  We will assume such thermal equilibrium  conditions to model  the contribution  to the total IR flux resulting from the known non-measured IR transitions (out of the range 10-30 $\mu$m) for which molar absorptivity is available  \citep{I-C22}  and apply an upward correction to  the total flux for each molecular specie. As we have used full aperture extraction spectra, band  fluxes are converted into I $_{tot}$  dividing by the subtended area in the sky by the corresponding slit of  each module of the IRS spectrograph.

The third parameter needed to infer column densities is the amino acid UV absorption cross section, $\sigma$$_{UV}$.  The aromatic amino acids contain conjugated aromatic rings and therefore are very efficient absorbing light in the UV range.  UV absorption cross sections  can be computed from laboratory measurements of molar absorption coefficients $\epsilon$ ($\lambda$) using the relation  $\sigma$($\lambda$)= 1000 $\epsilon$($\lambda$)/(N$_A$ log (e)) where N$_ A$ is the Avogadro number, and log (e) is the decimal logarithm of the Euler number. Given the expected stellar radiation field the most relevant wavelength range for determining effective UV absorption cross sections is from 190 to 230 nm. This is determined by the radiation emitted by the most luminous stars in IC 348. The resulting effective molar extinction coefficients,  $\epsilon_{eff}$ are 10850, 5640, 3530, 90 and 50 mol $^{-1}$ cm $^{-1}$,  for tryptophan, tyrosine, phenylalanine, isoleucine and glycine, respectively. Adopting these values, we obtain for the corresponding UV absorption cross sections 4, 2, 1, 0.03 and 0.02 x 10$^{-17}$cm $^{2}$, for tryptophan, tyrosine, phenylalanine, isoleucine and glycine, respectively.  The resulting column densities for the  aromatic amino acids are then: phenylalanine n(Phe)= 1 x 10 $^{11}$ cm $^{-2}$, tyrosine n(Tyr)= 0.8 x 10$^{11}$ cm $^{-2}$ , and tryptophan n(Trp)= 0.6 x 10 $^{11}$ cm $^{-2}$.  For  Isoleucine  we find a much higher column density n(Ile)= 2 x 10 $^{12}$ cm $^{-2}$. For glycine, the  column density appears to be the highest  n(Gly) = 9 x 10$^{12}$ cm $^{-2}$. 

Glycine is the most abundant amino acid in any type of carbonaceous chondrites with abundances of order 5-6 ppm in  CM2 types and  200 ppm in CR2 types, the meteorites with a higher abundance of amino acids. In both types,  isoleucine, phenylalanine and tyrosine are systematically  detected although  at lower concentrations \citep{Cobb14},  of order 1.0,  0.8 and 0.9 ppm, respectively, for  CM2 types.  In CR2 types, isoleucine is  more abundant with values of order 30 ppm, followed by phenylalanine 20 ppm, while tyrosine is at the level 0.7 ppm.   Among the amino acids studied in this work, glycine and isoleucine are the most abundant in CR2 meteorites. They  appear to be also the most abundant  in the ISM of IC 348.  On the other hand tryptophan, which we find to be  the less abundant of the five studied amino acids in Perseus, interestingly  it is  undetected in meteorites. Another relevant comparison with meteorites is the fraction of gas phase of carbon locked in amino acids.  The results are f$_C$ = 0.08, 0.09, 0.1, 1.8 and 2.4 x 10$^{-6}$  for tryptophan, tyrosine, phenylalanine, isoleucine and glycine, respectively. Remarkably, these values for isoleucine and glycine are very close to the  3 parts per million versus carbon reported in several types of carbonaceous chondrites.  
The emission features assigned to amino acids in the  ISM spectrum of IC 348 are also present in the combined  ISM spectrum from star-forming regions which also displays similar relative intensities. This suggests that amino acids may be  widely distributed across  the Galaxy. Future JWST observations at much higher S/N and spectral resolution  will show if this is the case.  

\section{CONCLUSIONS}
We have conducted a search  for mid-IR (10-30 $\mu$m) transitions of five amino acids (tryptophan, tyrosine, phenylalanine, glycine and isoleucine,) in the  Spitzer spectrum obtained averaging observations from various  ISM locations in the core of the IC 348 star-forming region. Each of the strongest laboratory bands of these amino acids have a counterpart emission feature in this spectrum. While reliable detections require better signal to noise and higher resolution  spectra (which can be obtained with JWST), from the tentative assignments proposed here,  preliminary column densities have been obtained. Interestingly, we find the relative abundance pattern of these molecules in the ISM of IC 348 would be similar to that  in meteorites, with glycine and isoleucine being much more abundant than any of the other  three aromatic amino acids. 
If, as suggested by these preliminary findings,  amino acids are present in the ISM of star-forming regions, they could also be part of the inventory of organic molecules in proto-planetary disks. Searches in protostars and pro-toplanetary disks  in Perseus and other molecular cloud complexes are worthwhile  as they  should  provide  valuable insight on  the delivery of complex organics  by meteoritic and cometary material to planets in early stages of formation and,  ultimately, on the  processes relevant  to the  origen of  life on Earth.
 
\begin{acknowledgements}
We acknowledge the support of the Gobierno de Canarias (Spain) via project P/302125GOB-CAN. We are indebted to Prof. F. Cataldo for his work on  characterization of amino acids and to the CASSIS team for the use of their public spectroscopic database.
\end{acknowledgements}

\end{document}